\pdfoutput=1
\documentclass[global,twocolumn,final]{svjour}

\usepackage{graphics}
\usepackage[usenames,dvipsnames,svgnames,table]{xcolor}
\usepackage{times,amsmath,amsfonts,bbm,braket}

\journalname{Applied Physics B}

\newcommand{\sbs}[1]{_{\text{#1}}}
\newcommand{\sps}[1]{^{\text{#1}}}

\begin{document}

\title{Overcoming lossy channel bounds using a single quantum repeater node}

\author{D. Luong\inst{1} \and L. Jiang\inst{2} \and J. Kim\inst{3} \and N. L\"{u}tkenhaus\inst{1}}

\institute{Department of Physics and Astronomy and Institute for Quantum Computing, \\ University of Waterloo, Waterloo, Ontario, Canada N2L 3G1 \and 
	Department of Applied Physics, Yale University, New Haven, CT 06511 USA \and 
	Electrical and Computer Engineering Department, Duke University, Durham, NC 27708, USA}

\date{\today{}}

\maketitle

\begin{abstract}
	We propose a scheme for performing quantum key distribution (QKD) which has the potential to beat schemes based on the direct transmission of photons between the communicating parties. In our proposal, the communicating parties exchange photons with two quantum memories placed between them. This is a very simple quantum repeater scheme and can be implemented with currently available technology. Ideally, its secret key rate scales as the square root of the transmittivity of the optical channel, which is superior to QKD schemes based on direct transmission because key rates for the latter scale at best linearly with transmittivity. Taking into account various imperfections in each component of our setup, we present parameter regimes in which our protocol outperforms protocols based on direct transmission.
\end{abstract}

\section{Introduction}
\label{sec:intro}

One of the outstanding problems of quantum key distribution (QKD) is the question of how to distribute key over arbitrarily long distances. The transmittivity of an optical channel decreases rapidly as the length of the channel grows (exponentially, in the case of fiber). This imposes a strong limit on the secret key rate achievable when photons are directly transmitted from Alice to Bob over long distances. Takeoka, Guha, and Wilde have shown that, when multi-mode signals are sent through a pure-loss bosonic channel with transmittivity $\eta\sbs{ch}$, the secret key rate can be no greater than $\log_2[(1+\eta\sbs{ch})/(1-\eta\sbs{ch})]$ bits per mode per channel use \cite{tgw14}. This is proportional to $\eta\sbs{ch}$ for small $\eta\sbs{ch}$, meaning that the key rate, too, decreases rapidly with distance. In order to improve this scaling behavior and achieve even a modest key rate at very long distances, it is necessary to look beyond direct transmission. 

One way to surpass the Takeoka-Guha-Wilde (TGW) bound is by using quantum repeaters \cite{guha14}. First described in \cite{briegel98a}, these are auxiliary quantum devices placed along the channel between the communicating parties, effectively breaking it up into multiple low-loss channels. A full repeater scheme might involve the use of many stations, each containing multiple qubits \cite{briegel98a,duan01a,sangouard11,jiang09,munro12,rpl09,fowler10,muralidharan14a}. These resource requirements are too demanding for such a scheme to be practical at present. Conceptually, then, quantum repeaters represent a path to transcending the limit imposed by the TGW bound, but their implementation remains a subject of intense research. No experiment has been performed that beats the TGW bound over any distance.

\begin{figure}
	\resizebox{\linewidth}{!}{\includegraphics{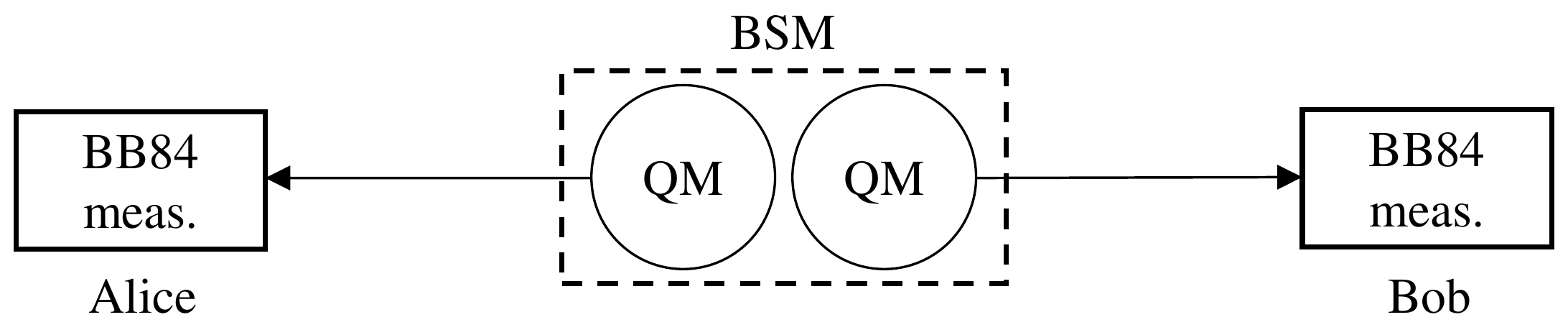}}
	\caption{Schematic of the proposed protocol. One QM sends entangled photons to Alice, the other to Bob. Once both parties successfully measure photons using BB84 measurements, a Bell measurement (BSM) is performed on the QMs.}
	\label{fig:protocol_diagram}
\end{figure}

In this paper, we propose a simplified scheme which has the potential to beat the TGW bound. Two parties perform QKD by measuring photons sent from a central station containing two quantum memories (Fig.~\ref{fig:protocol_diagram}). If the station is placed midway between the parties, each photon need only travel half the distance between them. Moreover, the presence of the memories means that the probability of one party successfully measuring a photon is independent of the success of the other party. Together, these imply that the secret key rate for our protocol is expected to scale as $\sqrt{\eta\sbs{ch}}$. Such scaling would be a fundamental improvement over any scheme relying on direct transmission, and gives it the potential to surpass the TGW bound. Our paper studies whether this scheme can beat the TGW bound in practice, taking into account experimental imperfections.

The protocol is basically a rudimentary quantum repeater scheme, but it can be implemented using currently available technology. Though not directly scalable, it represents a step towards the full implementation of quantum repeaters. It is similar in spirit to the protocol described in \cite{prml14}, except that in their proposal single photons or weak laser pulses are sent \emph{toward} the central station instead of being emitted \emph{from} the QMs. Our protocol thus simplifies the experimental requirements.

The remainder of this paper is organized as follows. Sec.~\ref{sec:protocol} describes our proposed protocol, and Sec.~\ref{sec:benchmarks} lists the benchmarks we will compare it to. Sec.~\ref{sec:model} describes how we mathematically model each component in our protocol; these models are assembled in Sec.~\ref{sec:keyrate} to produce an expression for the secret key rate of our protocol. In Sec.~\ref{sec:results}, we discuss the behavior of the key rate of our protocol as a function of the distance between the communicating parties. We also discuss variations of our protocol, and describe the parameter regimes required to beat the benchmarks listed in Sec.~\ref{sec:benchmarks}. In particular, we show that beating the TGW bound is within reach. Finally, we present some concluding remarks in Sec.~\ref{sec:conclusion}.

\section{Description of the protocol}
\label{sec:protocol}

The protocol we propose in this paper is illustrated in Fig.~\ref{fig:protocol_diagram}. It uses two quantum memories (QMs) in a central station placed between Alice and Bob, who wish to establish a secret key via QKD. We do not assume a particular implementation of the QMs, but we do require that each QM can be entangled with a single photon (as in, for example, ion-photon entanglement \cite{blinov04} or the DLCZ scheme \cite{duan01a}). The photonic degree of freedom used to encode qubits can be freely chosen; examples include polarization or time-bin encoding. We further assume that the two QMs can be jointly measured in the Bell basis, either by applying a CNOT gate and directly measuring them or by mapping the memory states onto photons and performing an optical Bell measurement. Alice and Bob are connected to the central station by lossy optical channels, and each have measurement apparatuses that allow them to measure incoming photons in one of two settings which correspond to mutually unbiased bases of the qubit subspace (as in BB84). We will call the bases $X$ and $Z$.

The procedure to produce one bit of raw key is as follows: 
\begin{enumerate}
	\item An entangled memory-photon state is prepared in one of the QMs and the photon sent to Alice, who performs a BB84 measurement on the photon. This is repeated until she successfully detects a photon.
	\item Same as the previous step, but with Bob and the other QM.
	\item A Bell measurement is performed on the two QMs and the result announced to Bob.
	\item If Bob measured in the $Z$ basis, he applies a bit flip to his BB84 measurement if the Bell measurement yielded $\ket{\Psi^+}$ or $\ket{\Psi^-}$. Similarly, if he measured in the $X$ basis, he applies a bit flip if the Bell measurement yielded $\ket{\Phi^-}$ or $\ket{\Psi^-}$.
\end{enumerate}
This procedure is repeated until a sufficient amount of raw key is obtained. The rest of the protocol is the same as in efficient BB84 \cite{lo00suba}.

The protocol described here admits of a few variations: the QMs could be \emph{simultaneously} or \emph{sequentially loaded} by performing steps 1 and 2 either at the same time or in sequence, and the position of the central station can be changed. In Sec.~\ref{sec:results}, we will explore the difference between simultaneous and sequential loading as well as the effect of changing the position of the central station.

\section{Benchmarks}
\label{sec:benchmarks}

In comparing our protocol to schemes based on the direct transmission of photons from Alice to Bob, the TGW bound \cite{tgw14} is the most stringent standard of comparison. We will, however, compare our protocol to other scenarios as well; this will make it easier to see how well it matches up to concrete schemes that can be performed in a lab. The direct transmission benchmarks with which we will compare our protocol are as follows:
\begin{enumerate}
	\item The TGW bound on the secret key rate per mode,
	\begin{equation}
	R\sbs{TGW} = \log_2 \! \left( \frac{1+\eta\sbs{ch}}{1-\eta\sbs{ch}} \right) \! ,
	\end{equation}
	where $\eta\sbs{ch}$ is the channel transmittivity. For small $\eta\sbs{ch}$, this reduces to $R\sbs{TGW} \approx{} (2/\ln 2)\eta\sbs{ch} \approx 2.89\eta\sbs{ch}$.
	\item BB84 with an ideal single-photon source and an ideal detector setup (no errors and no losses other than channel loss).
	\item BB84 with an ideal single-photon source and a realistic detector setup (nonzero misalignment error and dark counts, imperfect detector efficiency).
	\item Decoy-state BB84 with a laser and a realistic detector setup.
	\item BB84 using a quantum memory as a single photon source and a realistic detector setup.
\end{enumerate}
Throughout this paper, we will use the efficient variant of BB84 in which the $Z$ basis is measured much more frequently than the $X$ basis \cite{lo00suba}.

\begin{figure}
	\resizebox{\linewidth}{!}{\includegraphics{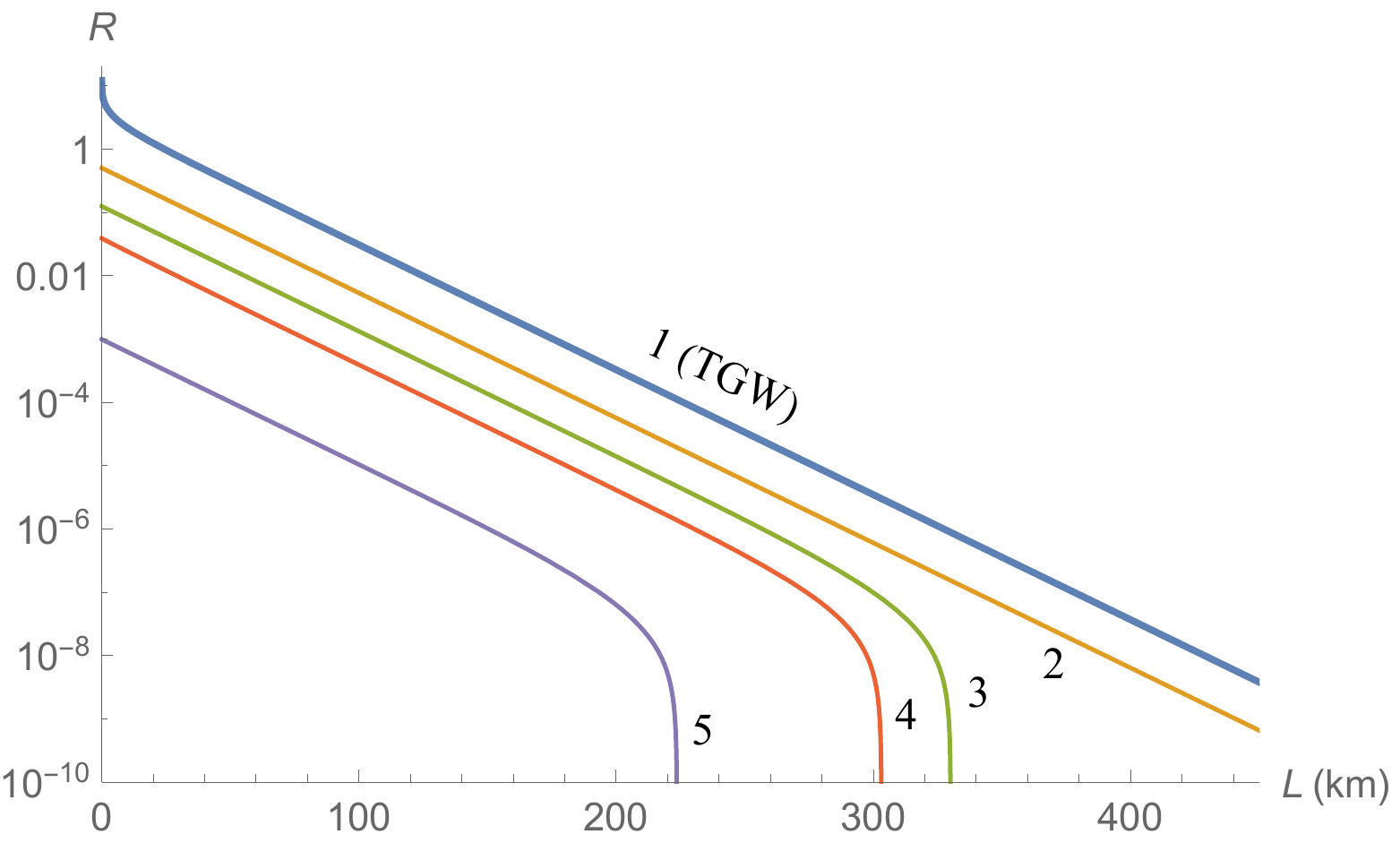}}
	\caption{Key rate per channel use per mode vs.\ distance for the benchmarks listed in Sec.~\ref{sec:benchmarks}. The thick curve corresponds to the TGW bound (benchmark 1). Parameter values are as given in Sec.~\ref{sec:results}.}
	\label{fig:benchmarks}
\end{figure}

The figure of merit to be considered is the key rate \emph{per mode}. Because BB84 requires two optical modes when implemented with the usual polarization or time-bin encoding, its key rate expression takes on a factor of 1/2. We may compare the key rate per mode of our protocol to those of the benchmarks above either on a per time unit or a per channel use basis. In this paper we will compare key rates per channel use only, though the ultimate goal is of course to compete on a per time unit basis. Any reference to ``key rates'' in the remainder of this paper, then, should be taken to mean ``key rates per mode per channel use''. Expressions for the key rates of benchmarks 2--5 are given in Appendix \ref{sec:appendix_BB84}.

Fig.~\ref{fig:benchmarks} shows plots of key rate per channel use as a function of the distance between Alice and Bob. Note that all of these benchmarks are proportional to $\eta\sbs{ch}$ (within certain limits, depending on the benchmark).

\section{Component modeling}
\label{sec:model}

In this section, we present a simple model of the behavior of each component in the setup described in Sec.~\ref{sec:protocol} in the absence of eavesdropping.

\subsection{Quantum memories}

In this paper, we consider QMs that are adequately described by the following model. A photon-memory entangled state can be generated in a QM with probability $\eta_p$; each attempt to do so requires a preparation time of $T_p$. When a photon is successfully generated, it is \emph{maximally entangled} with the QM; without loss of generality, we may take the initial memory-photon state to be the Bell state $\ket{\Phi^+}$. 

The memory-channel photon coupling efficiency is $\eta_c$. This includes not only the probability of a photon entering the optical channel, but the success probability of any process that occurs between the memory and the channel. For example, it contains the probability of successfully performing a wavelength conversion (if such is necessary).

The QM dephasing time is $T_2$. We will model dephasing using the following map \cite{rpl09}, which takes as input the initial state $\rho$ of the QM and returns the state of the QM after it has dephased for time $t$:
\begin{equation}
	\Gamma_t(\rho) := [1 - \lambda\sbs{dp}(t)] \rho + \lambda\sbs{dp}(t) Z\rho Z
\end{equation}
where
\begin{equation}
	\lambda\sbs{dp}(t) := \frac{1 - e^{-t/T_2}}{2}
\end{equation}
and $Z$ is the Pauli Z operator. In this model, the off-diagonal elements of $\rho$ go to zero as $t \to \infty$.

\subsection{Channels}

Alice and Bob are connected to the central station by optical channels of lengths $L_A$ and $L_B$ respectively; the length of the total channel is therefore $L = L_A + L_B$. The speed of light through these channels is $c$. The transmittivity of a channel of length $l$ is
\begin{equation}
	\eta\sbs{ch}(l) = e^{-l/L\sbs{att}}
\end{equation}
where $L\sbs{att}$ is the attenuation length.

The probability of error due to setup misalignment between Alice and the central station is $e_{mA}$. Setup misalignment rotates the qubit state of the photon; if we assume the rotation angle to be random and symmetrically distributed about 0, the initial memory-photon state $\ket{\Phi^+}$ becomes
\begin{equation}
	(1-e_{mA})\ket{\Phi^+}\!\bra{\Phi^+} + e_{mA}\ket{\Psi^-}\!\bra{\Psi^-}
\end{equation}
when the photon reaches a detector. This holds for Bob as well, with misalignment error $e_{mB}$.

\subsection{Detectors}

For their BB84 measurements, Alice and Bob each use a detector setup consisting of an optical element that can distinguish photonic qubit states in the $X$ and $Z$ bases (such as a polarizing beam splitter for polarization qubits) and two threshold detectors. We assume that they actively choose the basis in which to measure. Each detector has a dark count probability of $p_d$; each setup has efficiency $\eta_d$.

If a photon heading towards one of the setups is in the state $\rho$, the effect of dark counts can be mimicked by photons which are effectively in the modified state 
\begin{equation}
	\alpha(\eta) \rho + [1-\alpha(\eta)] \frac{\mathbbm{1}}{2},
\end{equation}
where
\begin{equation} \label{eq:alpha}
	\alpha(\eta) := \frac{\eta (1-p_d)}{1-(1-\eta)(1-p_d)^2}
\end{equation}
and $\eta$ is the probability that the photon reaches the detector setup. This assumes the use of a squashing map \cite{beaudry08a} which randomly assigns a measurement outcome to events in which both detectors click, reflected by $\rho$ being mapped into the maximally mixed state.

\subsection{Bell state measurement}

The probability of successfully performing a Bell state measurement (BSM) on the two QMs is $p\sbs{BSM}$.

We model errors in the BSM by applying the depolarizing channel 
\begin{equation}
	\Delta_{\lambda\sbs{BSM}}(\rho) = \lambda\sbs{BSM} \rho + (1-\lambda\sbs{BSM})\frac{\mathbbm{1}}{4}
\end{equation}
to the QMs before a perfect BSM. The parameter $\lambda\sbs{BSM}$ indicates how close the actual BSM is to an ideal BSM. When $\rho$ is pure, $\lambda\sbs{BSM}$ is related to the fidelity by
	\begin{equation}
		F\sbs{BSM} = \sqrt{\frac{3\lambda\sbs{BSM} + 1}{4}}.
	\end{equation}
	Because the fidelity is jointly concave, this is the minimum fidelity with which a BSM can be achieved.

\section{\label{sec:keyrate} Key rate analysis}

The secret key rate is lower bounded by \cite{lo00suba,scarani09a}
\begin{equation}
	R = \frac{Y}{2} [1 - h(e_X) - f h(e_Z)].
\end{equation}
Here, the yield $Y$ is the probability per channel use that Alice and Bob's measurements, as well as the BSM, were successful. $h(e)$ is the binary entropy function, $e_X$ and $e_Z$ are the quantum bit error rates (QBERs) between Alice and Bob in the $X$ and $Z$ bases, and $f$ is the error correction inefficiency. The factor of 1/2 comes from the fact that our protocol requires the use of two optical modes.

Because the total channel between Alice and Bob is divided in two by the central station and because the number of signals sent over each segment of the channel may in general be different, it is not immediately clear how to count channel uses. To be conservative, we define the number of channel uses required to produce one bit of raw key to be the \emph{greater} of the number of times Alice or Bob used their segments of the channel during the production of that bit. (Note that this is not the \emph{sum} of the number of times Alice and Bob used their segments of the channel, even in the case of sequential loading.)

\subsection{Yield}

The probability that a photon emitted from the central station is detected by Alice is 
\begin{equation}
\eta_A := \eta\sbs{tot} e^{-L_A/L\sbs{att}}.
	\end{equation}
where we have defined
\begin{equation}
	\eta\sbs{tot} := \eta_p \eta_c \eta_d.
\end{equation}
Due to the effect of dark counts, the probability that her detector clicks is
\begin{equation} \label{eq:etap}
	\eta_A' := 1 - (1-\eta_A)(1-p_d)^2.
\end{equation}

Let $N_A$ denote the number of photons that need to be sent to Alice so that her detector clicks once; it is a geometrically distributed random variable with success probability $\eta_A'$. Expressions similar to the above apply for Bob.

The average number of channel uses required for both Alice and Bob's detectors to click is $\mathbb{E}[\max(N_A,N_B)]$ where $\mathbb{E}$ is the expected value operator. The yield is therefore
\begin{align}
	Y &= \frac{p\sbs{BSM}}{\mathbb{E}[\max(N_A,N_B)]} \nonumber \\
	&= p\sbs{BSM} \! \left( \frac{1}{\eta'_A} + \frac{1}{\eta'_B} - \frac{1}{\eta'_A + \eta'_B - \eta'_A \eta'_B} \right)^{\!-1} \! .
\end{align}
The expectation value was evaluated in \cite{prml14}.

\subsection{Quantum bit error rates}

Taking into account all the parameters listed in Sec.~\ref{sec:model}, we find (in the absence of eavesdropping) that
\begin{align} 
\label{eq:e_X}
	e_X =\ &\lambda\sbs{BSM} \alpha(\eta_A) \alpha(\eta_B) [\varepsilon_m(1-\varepsilon\sbs{dp}) + (1-\varepsilon_m)\varepsilon\sbs{dp}] \nonumber
	\\ &+ \frac{1}{2} [1 - \lambda\sbs{BSM} \alpha(\eta_A) \alpha(\eta_B)] \\
\label{eq:e_Z}
	e_Z =\ &\lambda\sbs{BSM} \alpha(\eta_A) \alpha(\eta_B) \varepsilon_m + \frac{1}{2} [1 - \lambda\sbs{BSM} \alpha(\eta_A) \alpha(\eta_B)]
\end{align}
where
\begin{align} 
	\varepsilon_m &= e_{mA} (1-e_{mB}) + (1-e_{mA}) e_{mB} \\
\label{eq:dperror}
	\varepsilon\sbs{dp} &= \mathbb{E}\big[ \lambda\sbs{dp}(t_A)[1-\lambda\sbs{dp}(t_B)] + [1-\lambda\sbs{dp}(t_A)]\lambda\sbs{dp}(t_B) \big].
\end{align}
We may interpret $\varepsilon_m$ and $\varepsilon\sbs{dp}$ as the total misalignment and dephasing errors, respectively, between Alice and Bob. Here $t_A$ and $t_B$ are the times that Alice and Bob's QMs are left to dephase for.

At this point, we have fully determined $e_Z$ in terms of the parameters set out in Sec.~\ref{sec:model}. In order to evaluate $e_X$, we need only two more quantities: the dephasing time intervals $t_A$ and $t_B$. These are the subject of the following subsection.

\subsubsection{Dephasing}

Each time a QM emits a photon towards Alice, she must signal whether or not she successfully measured her photon before the QM prepares another one. This constrains the amount of time that elapses between photons to be at least
\begin{equation}\label{eq:rate}
	\tau_A = T_p + \frac{2 L_A}{c}.
\end{equation}
Similar remarks apply to Bob.

If it happens that $L_A \neq L_B$, then \eqref{eq:rate} allows the QMs to run at different rates. Throughout this paper, we will assume that each QM runs at the maximum rate allowed by \eqref{eq:rate}. It is possible to choose the rates to be the same, but we will not do so in this paper.

For both sequential and simultaneous loading, we may assume without loss of generality that Bob signals a successful measurement later than Alice does. The BSM is performed as soon as he does, so the QM that sends him photons dephases for a time
\begin{equation}
	t_B = \frac{2 L_B}{c}.
\end{equation}
The QM that sends photons to Alice dephases for a longer period of time because it must wait for Bob to make a successful measurement. If the QMs are sequentially loaded, Alice's QM dephases for
\begin{equation} \label{eq:tAseq}
	t_A\sps{seq} =  N_B \tau_B + \frac{2 L_A}{c}.
\end{equation}
If they are loaded simultaneously, then it dephases for
\begin{equation}
	t_A\sps{sim} = |N_B - N_A| \tau_B + \frac{2 L_A}{c}.
\end{equation}

In \eqref{eq:dperror}, because of the linearity of the operator $\mathbb{E}$, we need only evaluate $\mathbb{E}[e^{-t_A/T_2}]$. For sequential loading,
\begin{equation} \label{eq:expect_seq}
	\mathbb{E} \big[ e^{-t_A\sps{seq}/T_2} \big] = \frac{\eta'_B \exp\!\big( {-\frac{2 L_A}{c T_2}} \big)}{e^{\tau_B/T_2} + \eta'_B - 1}.
\end{equation}
For simultaneous loading, a result from \cite{prml14} gives
\begin{align}
	\mathbb{E} \big[ e^{-t_A\sps{sim}/T_2} \big] =\ &\frac{\eta'_A \eta'_B \exp\!\big( {-\frac{2 L_A}{c T_2}} \big)}{\eta'_A + \eta'_B -\eta'_A \eta'_B} \Bigg[ \frac{1}{1 - e^{-\tau_B/T_2}(1-\eta'_A)}  \nonumber \\ 
	&+ \frac{1}{1 - e^{-\tau_B/T_2}(1-\eta'_B)} - 1 \Bigg] \! .
\end{align}

\section{Results}
\label{sec:results}

Unless otherwise noted, the following parameter values were used for the results in this section. They are plausible values for an implementation of our protocol using trapped-ion quantum memories connected to Alice and Bob via optical fiber. A single ion fluorescence collection efficiency of 4.2\% has been demonstrated in \cite{streed11}, a trapped-ion qubit was measured to have a dephasing time of 2.5~s in \cite{olmschenk07}, and a two-qubit gate was used to entangle two ions with a fidelity of 99.3\% (corresponding to $\lambda\sbs{BSM} = 0.99$) in \cite{benhelm08}.
\begin{itemize}
	\item $\eta_p$ (preparation efficiency) $= 0.66$
	\item $T_p$ (preparation time) $= 2$ $\mu$s
	\item $\eta_c$ (photon-fiber coupling efficiency $\times$ wavelength conversion) $= 0.04 \times 0.3$
	\item $T_2$ (dephasing time) $= 1$ s
	\item $c$ (speed of light in optical fiber) $= 2 \times 10^8$ m/s 
	\item $L\sbs{att}$ (attenuation length) $= 22$ km
	\item $e_{mA}$ (misalignment error) $= e_{mB} = 0.01$
	\item $p_d$ (dark count probability per detector) $= 10^{-8}$
	\item $\eta_d$ (detector efficiency) $= 0.3$
	\item $p\sbs{BSM}$ (BSM success probability) $= 1$
	\item $\lambda\sbs{BSM}$ (BSM ideality parameter) $= 0.97$
	\item $f$ (error correction inefficiency) $= 1.16$
\end{itemize}
For decoy-state BB84 (benchmark 4), we will set the mean photon number of the signal states equal to 1. For the above numbers, we find that $\eta\sbs{tot} = \eta_p \eta_c \eta_d = 0.0024$.

\subsection{Protocol variations}

\subsubsection{Simultaneous vs.\ sequential loading}

For this comparison, the central station is located halfway between Alice and Bob.

\begin{figure}
	\resizebox{\linewidth}{!}{\includegraphics{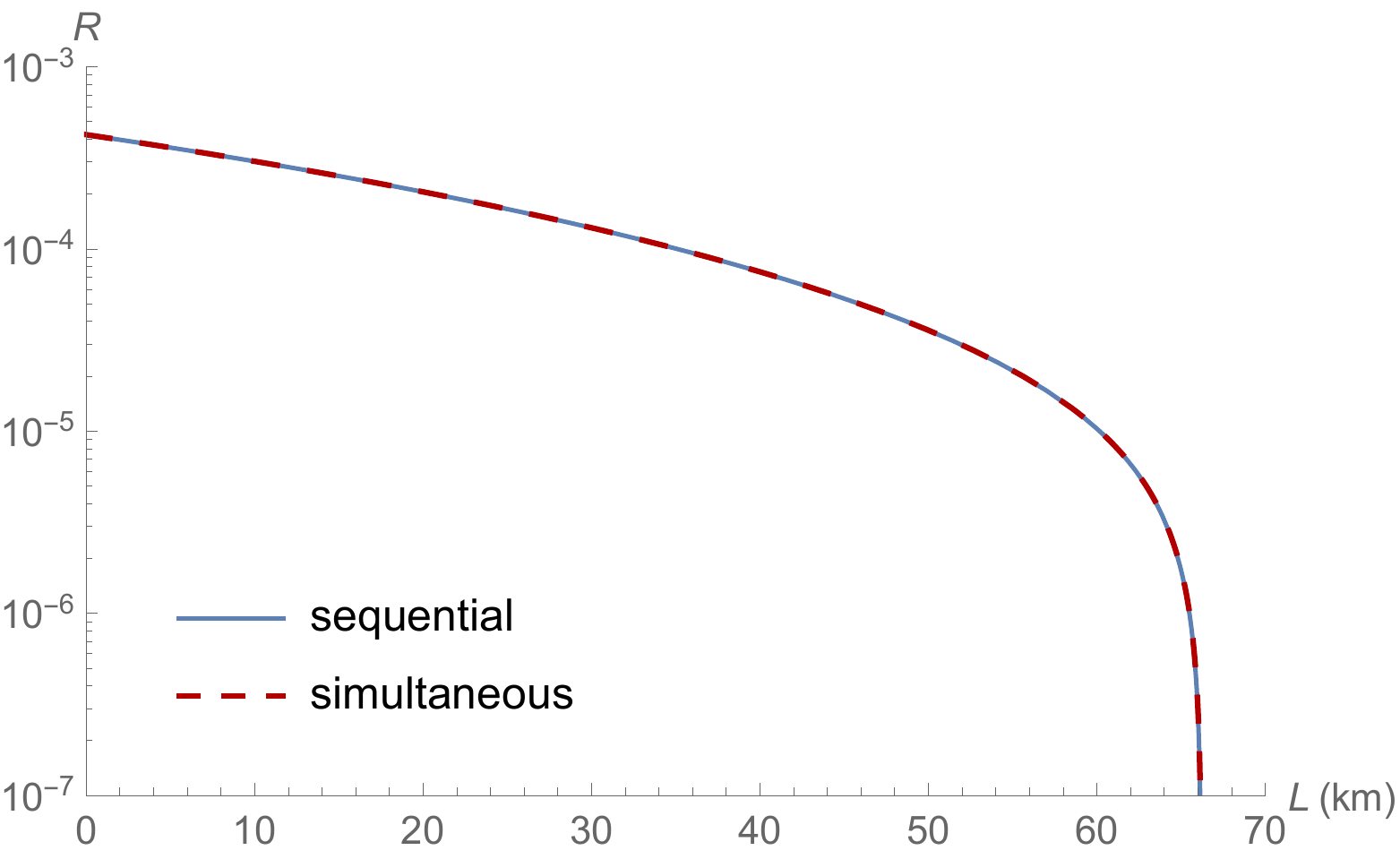}}
	\caption{\label{fig:sim_vs_seq} Key rate per channel use vs.\ distance for simultaneous and sequential loading. The two curves are virtually indistinguishable.}
\end{figure}

We have found, for the parameter values given above, that simultaneous and sequential loading of QMs in our protocol yield almost indistinguishable key rates per channel use over all values of $L$ for which the rates are nonzero (Fig.~\ref{fig:sim_vs_seq}). A rough comparison of the dephasing time intervals $t_A\sps{seq}$ and $t_A\sps{sim}$ suggests that this holds whenever $\tau_B/(T_2 \eta\sbs{tot})$ is small over all values of $L$ for which the key rates are nonzero. The parameters we have used are within this regime: $\tau_B/(T_2 \eta\sbs{tot}) = 0.140$ at $L = 66$~km. Outside of it, however, the difference can be dramatic: there are cases where the key rate is nonzero for simultaneous loading but not for sequential loading.

Because we will always be well within the parameter regime where simultaneous and sequential loading give nearly the same key rate, we will consider only sequential loading in the remainder of this paper.

\subsubsection{Optimization of central station position}

When the QMs are sequentially loaded, it need not be true that placing the central station halfway between Alice and Bob will yield the maximum key rate. This is because there is an inherent asymmetry in our protocol in this case: Bob only begins making measurements after Alice has finished hers.

\begin{figure}
	\resizebox{\linewidth}{!}{\includegraphics{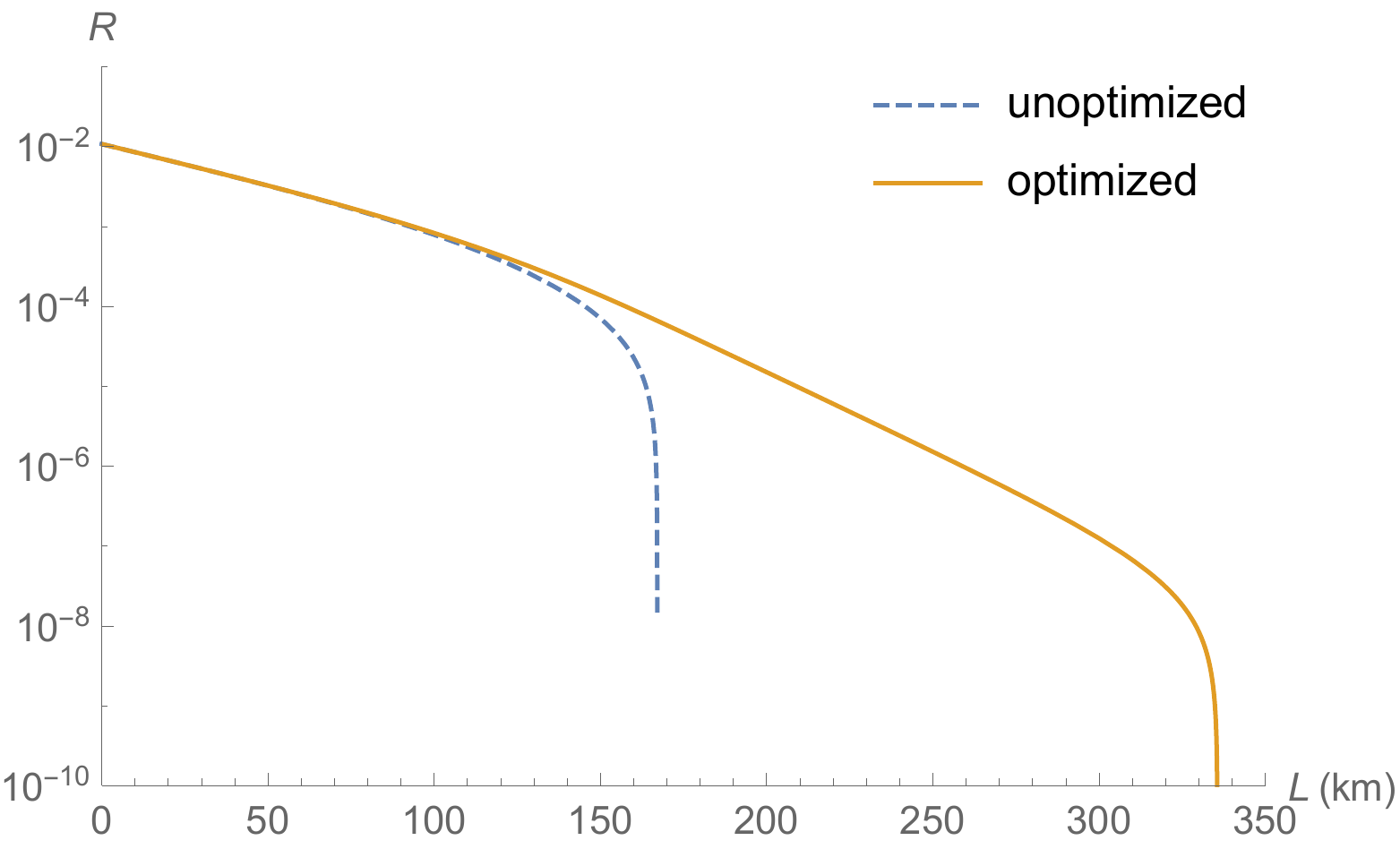}}
	\caption{\label{fig:opt_vs_unopt} Key rate vs.\ distance when the central station is at $L/2$ and when its position is optimized. ($\eta_c$ was increased to 0.3 to better show the features of the curves.) Near 150~km, the unoptimized key rate begins to drop to 0 and the optimized key rate transitions from $e^{-L/(2 L\sbs{att})}$ scaling to $e^{-L/L\sbs{att}}$ scaling.}
\end{figure}

Fig.~\ref{fig:opt_vs_unopt} shows the behavior of the key rate (per channel use) as a function of $L$ when the central station is placed at $L/2$ and when it is placed at the position that maximizes the key rate. For small $L$, both key rates are approximately the same, and scale proportionally to $\sqrt{\eta\sbs{ch}} = e^{-L/(2 L\sbs{att})}$. When $L$ becomes large enough for memory dephasing to become significant, the unoptimized key rate drops to zero. Around that same point, the optimized key rate transitions from $e^{-L/(2 L\sbs{att})}$ scaling to $\eta\sbs{ch} = e^{-L/L\sbs{att}}$ scaling---which is the same as for direct transmission---and continues thus until $L$ is so large that detector dark counts become significant, at which it too drops to zero.

\begin{figure}
	\resizebox{\linewidth}{!}{\includegraphics{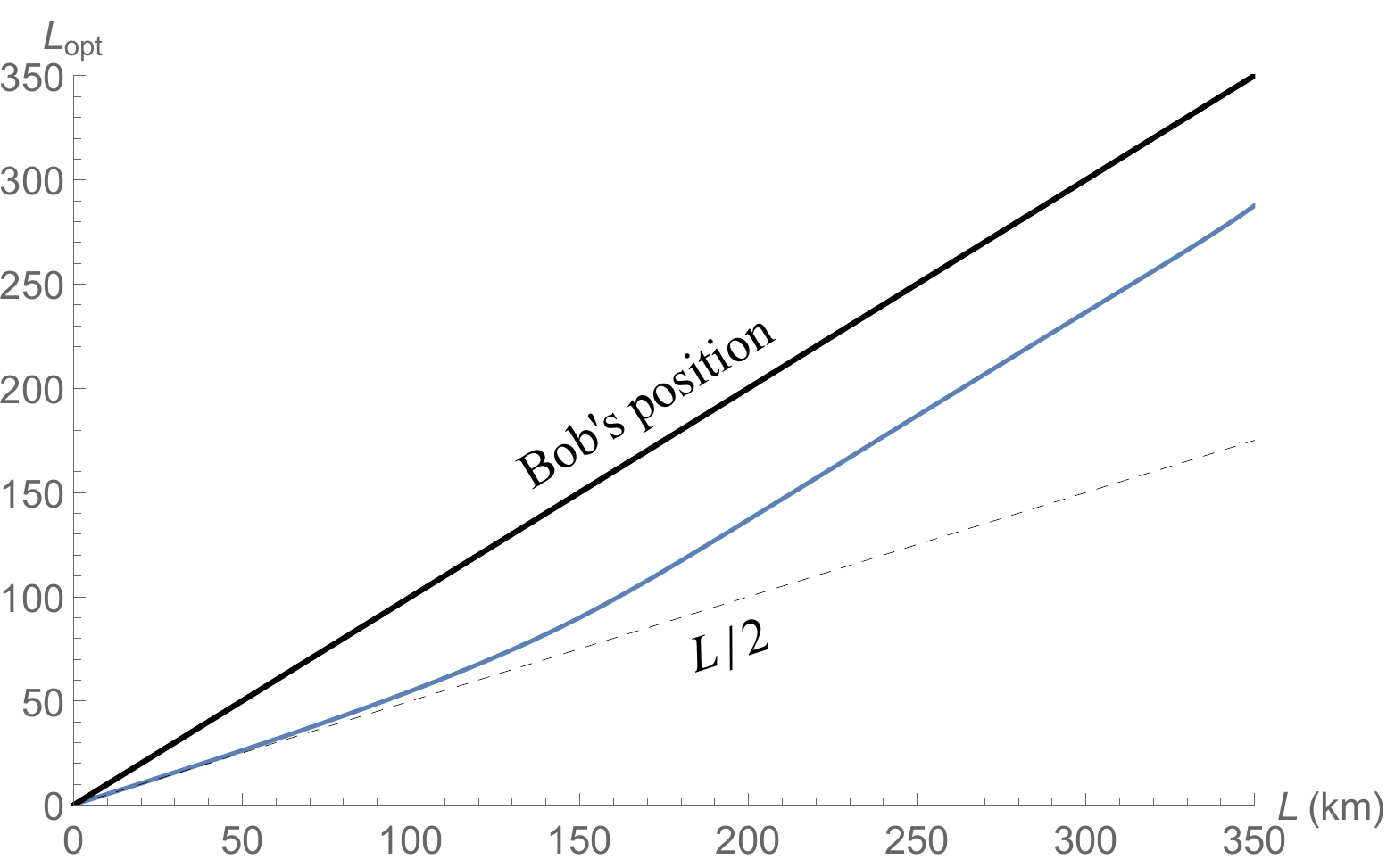}}
	\caption{\label{fig:opt_pos} Optimal central station position as a function of the total distance $L$. (Alice's position is taken to be at $L=0$.) Near where the scaling changes from $e^{-L/(2 L\sbs{att})}$ to $e^{-L/L\sbs{att}}$, around $L = 150$~km, the optimal position moves away from $L/2$ and remains a fixed distance away from Bob.}
\end{figure}

For greater insight into this behavior, consider Fig.~\ref{fig:opt_pos}, which shows the optimal central station position as a fraction of $L$ as $L$ is varied. For lower values of $L$, the station remains near the middle. Once dephasing becomes significant, the optimal position moves closer to Bob. This keeps dephasing errors low because Bob's link runs quicker, giving Alice's QM less time to dephase. At longer distances, the optimal position is a fixed distance away from Bob, just far enough away that the dephasing in Alice's QM does not overwhelm the system with errors. The price of suppressing dephasing errors in this way is that the key rate scales with the transmittivity of the longer link in the setup, so the key rate scaling is degraded.

\subsection{Beating direct transmission}

We are now in a position to determine the conditions under which our protocol can beat the direct transmission benchmarks listed in Sec.~\ref{sec:benchmarks}. First, note that at $L=0$ the performance of our protocol may be worse than that of the benchmarks because the central station introduces additional sources of loss. However, because the key rate for our protocol scales better with distance than the benchmark key rates when $L$ is not too large, \emph{crossover} with one or more of them is possible at some $L > 0$.

When the central station position is optimized, crossover can only occur in the $e^{-L/(2 L\sbs{att})}$ regime (excluding marginal cases)---that is, when the optimal position is near the midpoint between Alice and Bob. Equivalently, crossover can only occur when the unoptimized key rate is nonzero. For this reason, we will fix the central station at $L/2$ for the remainder of this section instead of optimizing its position. It is worth mentioning that crossover with a certain benchmark does not mean that our protocol beats it for \emph{all} $L$ beyond the crossover point; the interval over which our protocol is superior may be quite small. But optimizing the central station position can potentially increase the range of distances over which our protocol beats the benchmark compared to the leaving the station at $L/2$.

\begin{figure}
	\resizebox{\linewidth}{!}{\includegraphics{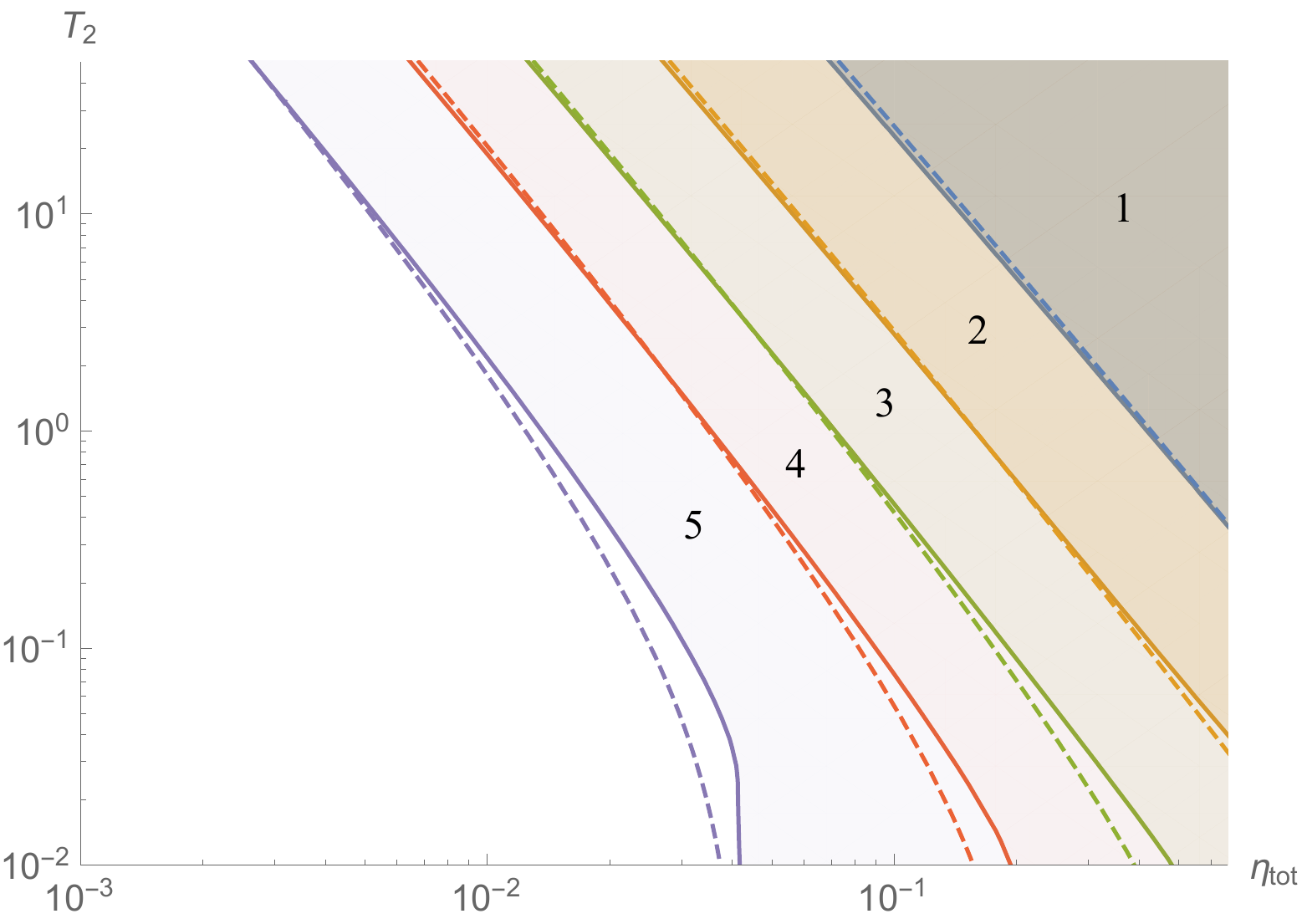}}
	\caption{\label{fig:reg_approx_exact} Regions in $\eta\sbs{tot}$-$T_2$ space where our protocol beats each of the benchmarks listed in Sec.~\ref{sec:benchmarks}, together with approximations of their boundaries obtained using \eqref{eq:approx} (dashed lines). For benchmark 5 (quantum memory as single photon source), we have fixed $\eta_c = 0.3 \times 0.3$.}
\end{figure}

We identify two parameters, the combined efficiency $\eta\sbs{tot}$ and the dephasing time $T_2$, which are crucial in determining whether crossover occurs with any of the benchmarks and which can be improved from the values given at the beginning of this section. For example, the photon-fiber coupling efficiency in $\eta_c$ could be pushed from 0.04 to as high as 0.3 \cite{kmk11} (leading to $\eta\sbs{tot} = 0.0178$), while a $T_2$ of 50~s has already been demonstrated \cite{lucas14}. Fig.~\ref{fig:reg_approx_exact} shows the regions in $\eta\sbs{tot}$-$T_2$ space in which we can beat each of the benchmarks. It is clear from the figure that we cannot beat any of the benchmarks with the parameters given at the beginning of the section, and that from our perspective, improving $\eta\sbs{tot}$ is more likely to result in crossover than improving $T_2$.

\begin{figure}
	\resizebox{\linewidth}{!}{\includegraphics{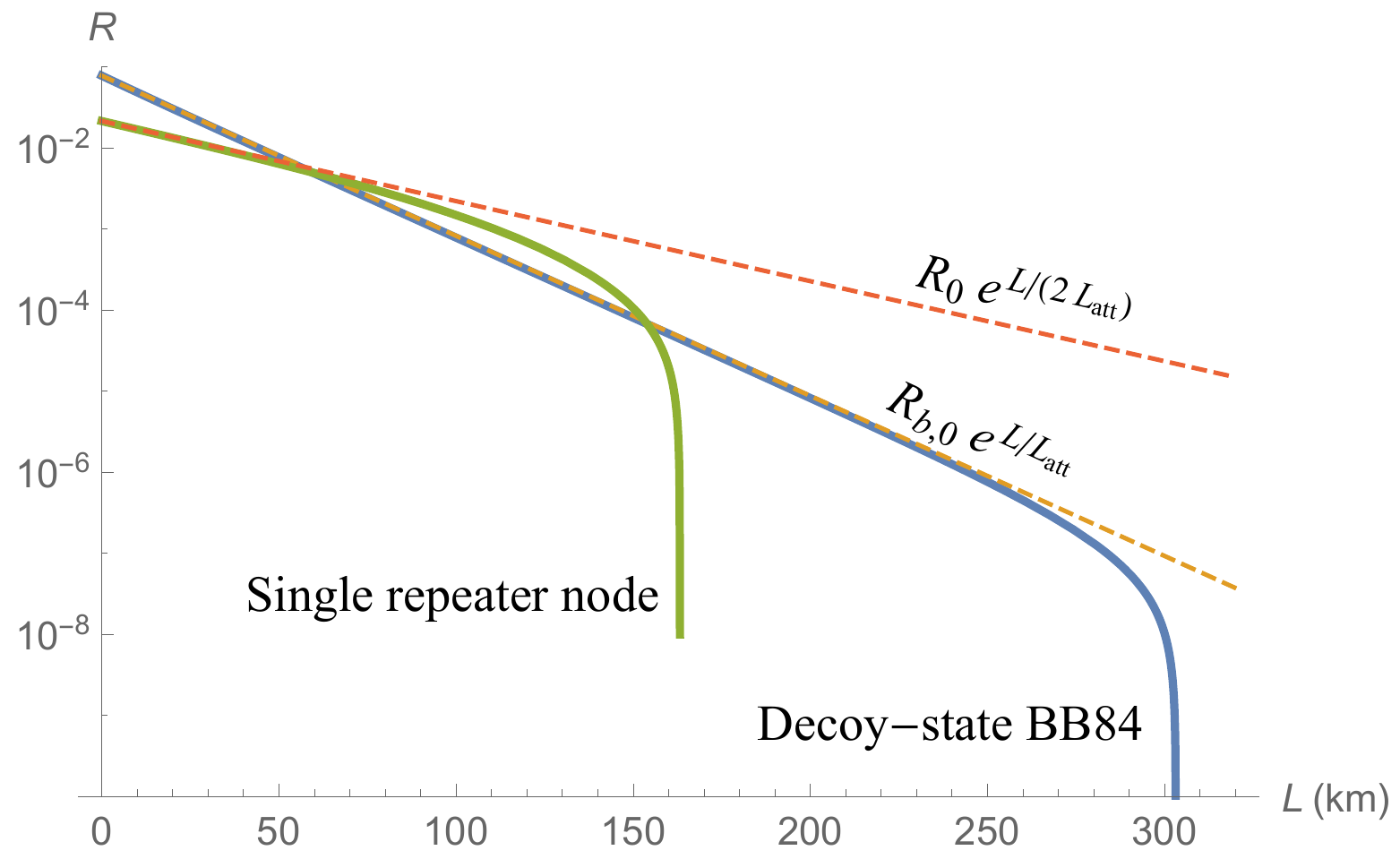}}
	\caption{\label{fig:approx_reg} Approximating the crossover point using the scaling behavior of the key rates. Note that the intersection point of the approximating curves coincides with the crossover point of the key rate curves, and that the intersection occurs before dephasing becomes significant and the key rate of our protocol goes to 0. (We have set $\eta_c = 0.3$.)}
\end{figure}

Each region may be explained in the following way. When $L$ is small enough for errors to be negligible, the key rate of our protocol is $R \approx R_0 e^{-L/(2 L\sbs{att})}$ while that of the benchmark of interest is $R_b \approx R_{b,0} e^{-L/L\sbs{att}}$, where $R_0$ and $R_{b,0}$ are the key rates at $L=0$ of our protocol and of the benchmark respectively.\footnote{This does not quite apply to the TGW bound, which goes to infinity as $L \to 0$. In this case, one must continue the $e^{-L/L\sbs{att}}$ behavior all the way to $L=0$, so that $R_{b,0} = 2/\ln 2$.} These curves intersect at a distance $L\sbs{int}$. If $L\sbs{int}$ is smaller than some characteristic distance $L\sbs{dp}$ beyond which dephasing becomes significant, then there is a crossover. The boundary of the crossover region corresponds to $L\sbs{int} = L\sbs{dp}$. These ideas are illustrated in Fig.~\ref{fig:approx_reg}.

Based on this explanation, we can derive an approximate formula for the boundary of the region in which crossover occurs with a given benchmark with key rate $R_b$:
\begin{equation} \label{eq:approx}
	T_2 = K \! \left[ \frac{Q T_p}{\eta\sbs{tot}^2} + \frac{2 L\sbs{att} \ln(Q/\eta\sbs{tot})}{c} \left( 1 + \frac{Q}{\eta\sbs{tot}^2} \right) \right] \! .
\end{equation}
Here
\begin{equation} \label{eq:quotient}
	Q = \frac{3 R_{b,0}}{2 R_0^{\eta\sbs{tot}=1}},
\end{equation}
$R_0^{\eta\sbs{tot}=1}$ denotes the key rate of our protocol when $L=0$ and $\eta\sbs{tot} = 1$, and $K$ is a fitting parameter characterizing how long the QMs must dephase for, as a fraction of $T_2$, before dephasing becomes significant. It needs to be chosen to fit the exact crossover region boundary; empirically, $K = 14$ gives a good fit. This approximation is valid when $T_p \ll T_2$ and $p_d \ll \eta\sbs{tot}^2/Q$. A derivation is given in Appendix \ref{sec:appendix_approx}.

The dashed lines in Fig.~\ref{fig:reg_approx_exact} are the boundary approximations given by \eqref{eq:approx}.

\subsubsection{Attenuation length; the high-loss limit}

Let us now consider the \emph{high-loss limit}, where the attenuation length $L\sbs{att}$ is very small. This limit is interesting in the context of hybrid quantum-classical networks. In passive optical networks, where multiple users are connected to a source, each user is effectively connected to the source via a high-loss channel. The limit is also applicable when the wavelength of the photons emitted by the QMs happens to be greatly attenuated by the optical channel.

\begin{figure}
	\resizebox{\linewidth}{!}{\includegraphics{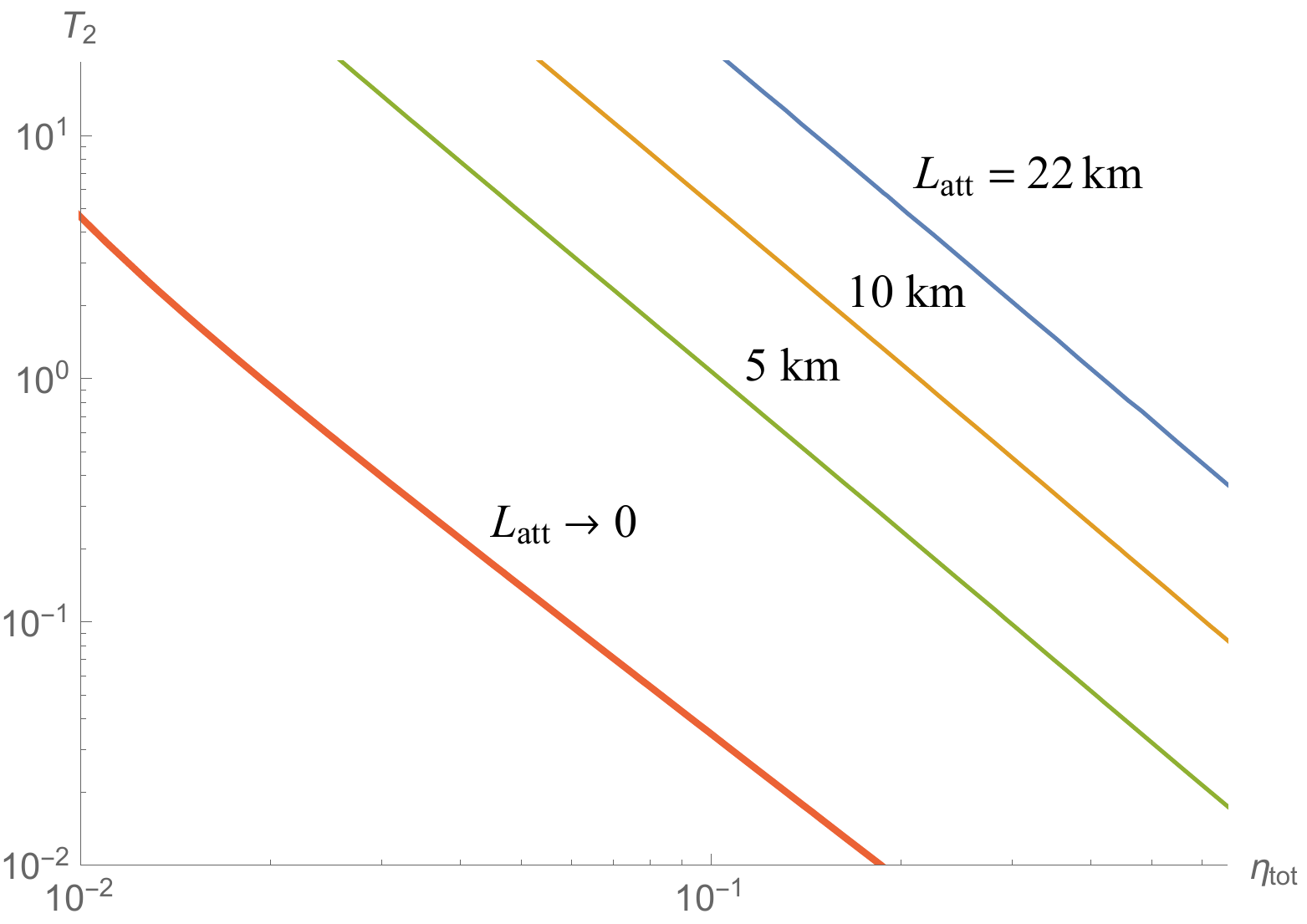}}
	\caption{Boundaries of regions in $\eta\sbs{tot}$-$T_2$ space where our protocol beats the TGW bound for various attenuation lengths.}
	\label{fig:attlen}
\end{figure}

The effect of reducing the attenuation length is to make it easier to beat the benchmarks, as shown in Fig.~\ref{fig:attlen} and predicted in \eqref{eq:approx}. This is because the photons cannot travel as far, so there is less dephasing. However, because of the nonzero preparation time $T_p$, beating the benchmarks is still nontrivial in the $L\sbs{att} \to 0$ limit. The high-loss limit thus represents a regime in which experimental requirements are relaxed, yet the benchmarks can still meaningfully be beaten.

\begin{figure}
	\resizebox{\linewidth}{!}{\includegraphics{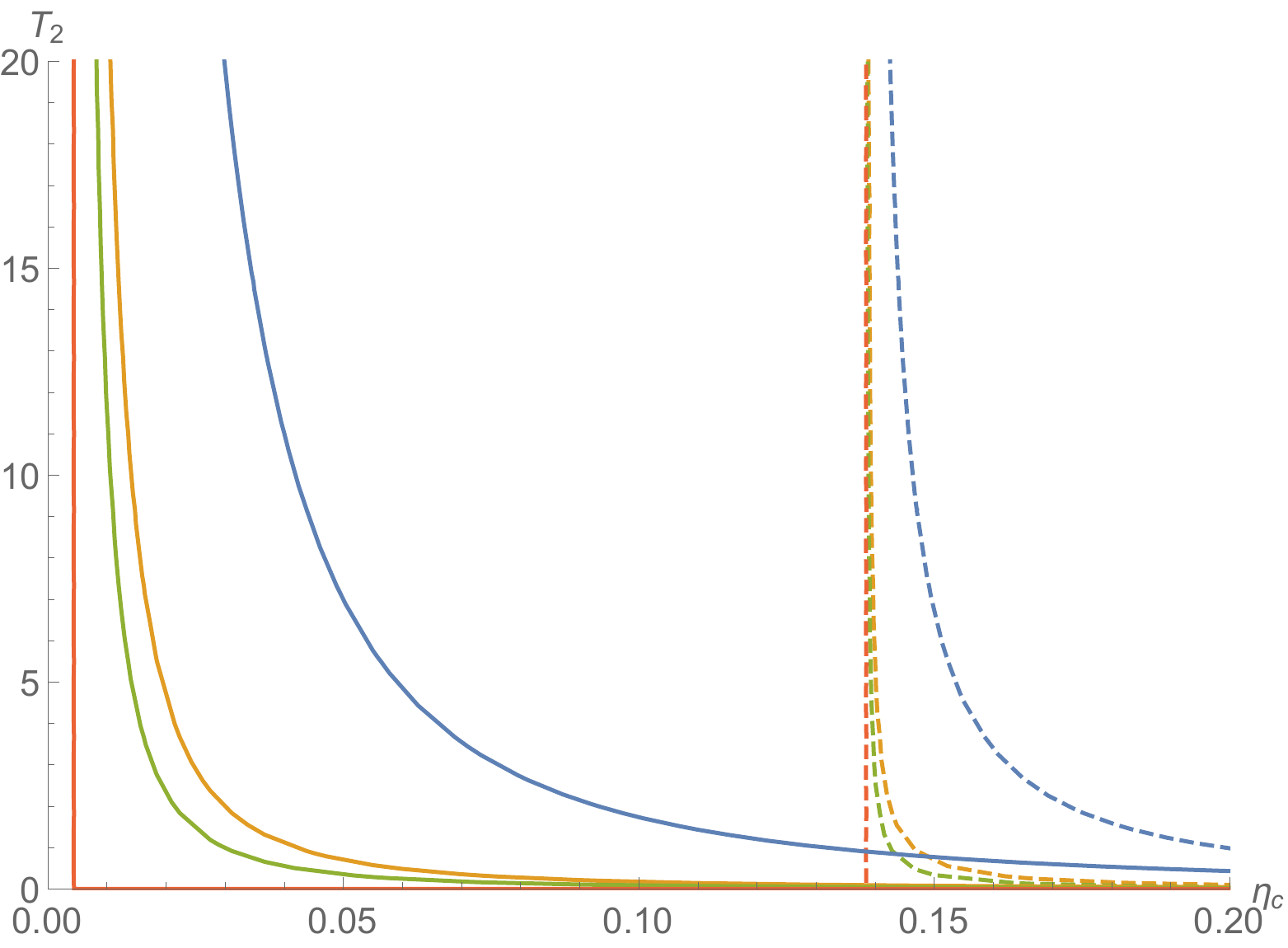}}
	\caption{\label{fig:tprep_dark} Boundaries of regions in $\eta\sbs{tot}$-$T_2$ space where our protocol beats the TGW bound in the limit $L\sbs{att} \to 0$. Solid lines indicate $p_d = 10^{-8}$, dashed lines $p_d = 10^{-5}$. Blue, orange, green, and red lines indicate $T_p = 100$, 10, 5, and 0 $\mu$s respectively.}
\end{figure}

Fig.~\ref{fig:tprep_dark} shows the effect of changing the preparation time $T_p$ and the dark count probability $p_d$ on the $\eta\sbs{tot}$-$T_2$ regions in which our protocol can beat the TGW bound. As expected from \eqref{eq:approx}, the benchmarks become easier to beat as $T_p$ goes down. (This is true whatever the value of $L\sbs{att}$.) We also see that when $T_p = 0$ and $L\sbs{att} \to 0$, they can be beat for any value of $T_2$. Because there is no dephasing at all in this case, $T_2$ plays no role in determining whether there is a crossover. 

Equation \eqref{eq:approx} suggests that when $T_p = 0$ and $L\sbs{att} \to 0$, crossover can happen for any value of $\eta\sbs{tot}$. However, Fig.~\ref{fig:tprep_dark} shows that crossover can happen only when $\eta\sbs{tot}$ is sufficiently large. There is no contradiction: when $\eta\sbs{tot}$ is too low, the condition $p_d \ll \eta\sbs{tot}^2/Q$ is violated and \eqref{eq:approx} no longer holds. It turns out that there is no crossover when $\eta\sbs{tot}$ is small because dark counts become significant. Using reasoning similar to that employed in deriving \eqref{eq:approx} (with the characteristic dephasing length $L\sbs{dp}$ replaced with a characteristic dark count length $L_d$), we can obtain the following approximation to the minimum $\eta\sbs{tot}$ necessary for our protocol to beat a given benchmark:
\begin{equation} \label{eq:min_coupling}
	\eta\sbs{tot}\sps{min} = \sqrt{\frac{Q p_d (2-p_d)(1-\xi)}{(1-p_d)(p_d+\xi-p_d\xi)}}.
\end{equation}
The quotient of key rates at zero distance, $Q$, is as defined in \eqref{eq:quotient}, and depends on the choice of benchmark. The fitting parameter $\xi$ is a measure of how much error due to dark counts our system can tolerate before the key rate drops to zero. For the parameter values given at the beginning of the section, $\xi = 0.012$ fits well. This equation is valid when $T_p \ll T_2$ and $p_d \ll \eta\sbs{tot}$. The derivation is in Appendix \ref{sec:appendix_approx}.

\section{\label{sec:conclusion} Conclusion}

In this paper, we have analyzed a QKD protocol in which Alice and Bob exchange signals with a central station consisting of two quantum memories: a rudimentary quantum repeater node. We have also introduced a number of benchmarks to which our protocol can be compared, the most important of them being the Takeoka-Guha-Wilde bound on the secret key rate. We showed that our protocol can, in principle, beat the benchmarks because of its improved rate-vs.-distance scaling: the key rate of all protocols relying on direct transmission between Alice and Bob scales at best with $e^{-L/L\sbs{att}}$, while our protocol scales as $e^{-L/(2 L\sbs{att})}$. In effect, our protocol doubles the attenuation length. Finally, we explored the conditions under which we can beat the benchmarks in practice.

Because our protocol uses only one intermediate station with only two memories, we do not obtain the full scalability that quantum repeaters can offer us in theory. However, by the same token, it is feasible to implement using currently available technology while still exhibiting the rate improvement of a full quantum repeater scheme and the ability to beat the TGW bound. Beating the bound would, in and of itself, be a fundamental experimental achievement---an achievement which we have shown to be within reach, particularly in the high-loss limit. Our protocol, then, is a first step towards the experimental implementation of quantum repeaters.

\begin{acknowledgement}
	We would like to thank Mohsen Razavi for insightful discussions, including a hint regarding the relevance of the high-loss limit to quantum networks. We also thank Ryo Namiki for many fruitful discussions. This work has been supported by an NSERC Discovery Grant, the DARPA QUINESS program, and Industry Canada.
\end{acknowledgement}

\appendix

\section{Benchmark key rates}
\label{sec:appendix_BB84}

In this appendix, we give expressions for the secret key rates of benchmarks 2--5 in Sec.~\ref{sec:benchmarks}. For this purpose we model the observables in an experimental demonstration operating normally---that is, in the absence of eavesdropping activity.

For benchmarks 2, 3, and 5, Alice transmits single photons to Bob. In this case the efficient BB84 key rate per mode is lower bounded by \cite{lo00suba,scarani09a}
\begin{equation}
	R = \frac{Y_1}{2} [1 - h(e_1) - f h(e_1)]
\end{equation}
where
\begin{align}
\begin{split}
	\eta &:= \eta\sbs{tot} e^{-L/L\sbs{att}} \\
	Y_1 &:= 1-(1-\eta)(1-p_d)^2 \\
	e_1 Y_1 &:= Y_1/2 - (1/2 - e_m) \eta (1 - p_d).
\end{split}
\end{align}
Here $Y_1$ and $e_1$ are the yield and QBER for single photons, $f$ is the error correction inefficiency, $L$ is the length of the optical channel between Alice and Bob, and $e_m$ is the setup misalignment error probability. The other variables are as defined in Sec.~\ref{sec:model}. The factor of 1/2 comes from the fact that BB84 uses two optical modes.

For an ideal single photon source (benchmark 3), $\eta_p = \eta_c = 1$. For an ideal detector setup (benchmark 2), $\eta_d = 1$ and $p_d = e_m = 0$. This amounts to setting $e_1 = 0$ and results in $R = e^{-L/L\sbs{att}}/2 = \eta\sbs{ch}/2$.

The key rate for decoy-state BB84 with a laser (benchmark 4) is \cite{lo05a}
\begin{equation}
R\sbs{decoy} = \frac{1}{2}\big( Q_1[1-h(e_1)] -f Q_\mu h(E_\mu) \big)
\end{equation}
where
\begin{align}
\begin{split}
	Q_1 &:= Y_1 \mu e^{-\mu} \\
	Q_\mu &:= 1 - e^{-\eta\mu} (1-p_d)^2 \\
	E_\mu Q_\mu &:= Q_\mu/2 - (1/2 - e_m)(1 - e^{-\eta\mu})(1 - p_d).
\end{split}
\end{align}
Here $\mu$ is the average photon number for signal states; $Y_1$, $e_1$, $f$, and $e_m$ are as defined above.

\section{Approximation of crossover regions in $\eta\sbs{tot}$-$T_2$ space}
\label{sec:appendix_approx}

Throughout this appendix, we will assume that the QMs are loaded sequentially and that the central station is at $L/2$. Let $R_b$ be the key rate for the benchmark whose crossover region we wish to approximate.

We will first derive \eqref{eq:approx}. As outlined in the discussion leading up to that equation, our approach is to equate the intersection of the curves $R_0 e^{-L/(2 L\sbs{att})}$ and $R_{b,0} e^{-L/L\sbs{att}}$ with some characteristic dephasing length $L\sbs{dp}$ in order to find the boundary of the crossover region. ($R_0$ and $R_{b,0}$ are the key rates at $L=0$ of our protocol and of the benchmark, respectively.)

The first step is to find conditions under which
\begin{equation} \label{eq:approx_R_at_zero}
	R_0 \propto \eta\sbs{tot} R_0^{\eta\sbs{tot}=1}.
\end{equation}
If $p_d$ is small and $T_p \ll T_2$, then $e_X$ and $e_Z$ are approximately independent of $\eta\sbs{tot}$---see \eqref{eq:alpha} and \eqref{eq:expect_seq}---and $R_0$ only depends on $\eta\sbs{tot}$ through $Y$. If we further assume that $\eta'_A \approx \eta_A$, then
\begin{equation}
	Y = p\sbs{BSM} \frac{\eta\sbs{tot} (2 - \eta\sbs{tot})}{3 - 2 \eta\sbs{tot}} \approx \frac{2}{3} p\sbs{BSM} \eta\sbs{tot}.
\end{equation}
to first order in $\eta\sbs{tot}$. These conditions are therefore sufficient for the approximation in \eqref{eq:approx_R_at_zero} to hold, with proportionality constant 2/3. 

Given this fact, the intersection of the two curves is at
\begin{equation}
	L\sbs{int} = 2 L\sbs{att} \ln\!\left(\frac{Q}{\eta\sbs{tot}}\right)
\end{equation}
where $Q$ is defined in \eqref{eq:quotient}. Note that $T_p \ll T_2$ implies that $Q$ is independent of $T_2$.

We now derive a characteristic dephasing length by determining the distance at which Alice's QM dephases for a significant fraction of $T_2$. (Recall that Alice's QM always dephases longer than Bob's.) That is, we put
\begin{equation} \label{eq:avg_deph}
	\frac{T_2}{K} = \mathbb{E}(t_A\sps{seq}) = \frac{L\sbs{dp}}{c} + \frac{T_p + L\sbs{dp}/c}{\eta\sbs{tot} e^{-L\sbs{dp}/(2L\sbs{att})}}
\end{equation}
where we have again used $\eta'_A \approx \eta_A$. The fitting parameter $K$ defines the fraction of $T_2$ at which dephasing becomes significant.

Equation \eqref{eq:avg_deph} cannot be solved for the dephasing length $L\sbs{dp}$ using elementary functions, but this is unnecessary: to find the crossover boundary, we need only substitute $L\sbs{int}$ for $L\sbs{dp}$. After a minor rearrangement of terms, this yields \eqref{eq:approx}.

It may appear that a small $p_d$ implies that $\eta'_A \approx \eta_A$. It is true that $p_d \ll 1$ implies $|\eta'_A - \eta_A| \ll 1$, but since $\eta_A \ll 1$ and $\eta_A' \ll 1$ in general, this is not strong enough to meaningfully say that $\eta'_A \approx \eta_A$. We require instead that $|\eta'_A - \eta_A|/\eta_A \ll 1$. Moreover, because we have used $\eta'_A \approx \eta_A$ in deriving \eqref{eq:avg_deph}, we require this to hold for all $L$ up to $L\sbs{dp}$---or, equivalently, up to $L\sbs{int}$. By manipulating \eqref{eq:etap}, we can write
\begin{equation}
	\frac{|\eta'_A - \eta_A|}{\eta_A} = \left( \frac{1}{\eta_A} - 1 \right) \! (2p_d - p_d^2) \approx \left( \frac{1}{\eta_A} - 1 \right) \! p_d.
\end{equation}
If $\eta_A$ is close to 1, then $(1/\eta_A - 1) p_d$ is already small and the approximation holds. If $\eta_A \ll 1$, then $(1/\eta_A - 1)p_d \approx p_d/\eta_A$, which is small for all $L$ up to $L\sbs{int}$ when $p_d \ll \eta\sbs{tot}^2/Q$. This condition, then, together with $T_p \ll T_2$, guarantees the validity of \eqref{eq:approx}.

Let us now derive \eqref{eq:min_coupling}. This time, we will compare $L\sbs{int}$ with a length $L_d$ at which errors due to dark counts become significant.

The error due to dark counts is related to $\alpha(\eta_A)$, defined in \eqref{eq:etap}. We will put $1-\xi = \alpha(\eta_A)$ where $\xi$ is a parameter indicating the amount of error the system can tolerate due to dark counts. Rearranging this equation, we obtain
\begin{equation}
	L_d = 2 L\sbs{att} \ln \! \left( \frac{\eta\sbs{tot} (1-p_d)(p_d+\xi-p_d\xi)}{p_d (2-p_d)(1-\xi)} \right) \! .
\end{equation}
By equating $L_d$ and $L\sbs{int}$, we obtain \eqref{eq:min_coupling}.

In deriving this equation, we have made no assumptions beyond those required for \eqref{eq:approx_R_at_zero}. In particular, we do not require $\eta'_A \approx \eta_A$ for all $L$ up to $L\sbs{int}$, but only at $L = 0$. This means that the condition on $p_d$ is less strict: $p_d \ll \eta\sbs{tot}$.

Finally, we note that the condition $p_d \ll \eta\sbs{tot}^2/Q$, required for \eqref{eq:approx}, can be obtained from a linearization of the square of \eqref{eq:min_coupling}.

\bibliographystyle{h-physrev}
\bibliography{qit_2015_03_24}

\begin{thebibliography}{10}

\bibitem{tgw14}
M.~Takeoka, S.~Guha, and M.~M. Wilde,
\newblock IEEE Transactions on Information Theory {\bf 60}, 4987 (2014).

\bibitem{guha14}
S.~Guha {\em et~al.},
\newblock Rate-loss analysis of an efficient quantum repeater architecture,
  2014, arXiv:1404.7183.

\bibitem{briegel98a}
H.~J. Briegel, W.~D\"ur, J.~I. Cirac, and P.~Zoller,
\newblock Phys. Rev. Lett. {\bf 81}, 5932 (1998).

\bibitem{duan01a}
L.~M. Duan, M.~D. Lukin, J.~I. Cirac, and P.~Zoller,
\newblock Nature {\bf 414}, 413 (2001).

\bibitem{sangouard11}
N.~Sangouard, C.~Simon, H.~de~Riedmatten, and N.~Gisin,
\newblock Rev. Mod. Phys. {\bf 83}, 33 (2011).

\bibitem{jiang09}
L.~Jiang {\em et~al.},
\newblock Phys. Rev. A {\bf 79}, 032325 (2009).

\bibitem{munro12}
W.~Munro, A.~Stephens, S.~Devitt, K.~Harrison, and K.~Nemoto,
\newblock Nature Photonics {\bf 6}, 777 (2012).

\bibitem{rpl09}
M.~Razavi, M.~Piani, and N.~L\"utkenhaus,
\newblock Phys. Rev. A {\bf 80}, 032301 (2009).

\bibitem{fowler10}
A.~G. Fowler {\em et~al.},
\newblock Phys. Rev. Lett. {\bf 104}, 180503 (2010).

\bibitem{muralidharan14a}
S.~Muralidharan, J.~Kim, N.~L\"utkenhaus, M.~D. Lukin, and L.~Jiang,
\newblock Phys. Rev. Lett. {\bf 112}, 250501 (2014).

\bibitem{prml14}
C.~Panayi, M.~Razavi, X.~Ma, and N.~L{\"u}tkenhaus,
\newblock New Journal of Physics {\bf 16}, 043005 (2014).

\bibitem{blinov04}
B.~Blinov, D.~Moehring, L.-M. Duan, and C.~Monroe,
\newblock Nature {\bf 428}, 153 (2004).

\bibitem{lo00suba}
H.~K. Lo, F.~Chau, and M.~Ardehali,
\newblock J. Cryptology {\bf 18}, 133 (2005).

\bibitem{beaudry08a}
N.~J. Beaudry, T.~Moroder, and N.~L\"utkenhaus,
\newblock Phys. Rev. Lett. {\bf 101}, 093601 (2008).

\bibitem{scarani09a}
V.~Scarani {\em et~al.},
\newblock Rev. Mod. Phys. {\bf 81}, 1301 (2009).

\bibitem{streed11}
E.~W. Streed, B.~G. Norton, A.~Jechow, T.~J. Weinhold, and D.~Kielpinski,
\newblock Phys. Rev. Lett. {\bf 106}, 010502 (2011).

\bibitem{olmschenk07}
S.~Olmschenk {\em et~al.},
\newblock Phys. Rev. A {\bf 76}, 052314 (2007).

\bibitem{benhelm08}
J.~Benhelm, G.~Kirchmair, C.~F. Roos, and R.~Blatt,
\newblock Nature Physics {\bf 4}, 463 (2008).

\bibitem{kmk11}
T.~Kim, P.~Maunz, and J.~Kim,
\newblock Phys. Rev. A {\bf 84}, 063423 (2011).

\bibitem{lucas14}
T.~P. Harty {\em et~al.},
\newblock Phys. Rev. Lett. {\bf 113}, 220501 (2014).

\bibitem{lo05a}
H.-K. Lo, X.~Ma, and K.~Chen,
\newblock Phys. Rev. Lett. {\bf 94}, 230504 (2005).

\end{thebibliography}

\end{document}